\documentclass[a4paper]{jpconf}
\usepackage{amsmath}
\usepackage{graphicx}
\usepackage{latexsym}  
\usepackage{amsfonts}  
\usepackage{amssymb} 
\usepackage{array}
\usepackage{color,mathtools}
\usepackage{dsfont,xcolor}
\usepackage{url}
\usepackage{booktabs}
\usepackage[square,sort&compress,numbers]{natbib}
\usepackage{verbatim}
\usepackage[utf8x]{inputenc}
\usepackage{feynmp-auto}

\newcommand{\be}{\begin{equation}}  
\newcommand{\ee}{\end{equation}}
\newcommand{\beq}{\begin{eqnarray}}  
\newcommand{\eeq}{\end{eqnarray}}  

\begin{document}

\title{Development of non-equilibrium Green's functions for use with full interaction in complex systems}

\author{Richard A Lynn$^{1,2}$, Robert van Leeuwen$^{1,2}$}

\address{$^1$ Department of Physics, Nanoscience Center, FIN 40014,
University of Jyv\"askyl\"a, Jyv\"askyl\"a, Finland}
\address{$^2$ European Theoretical Spectroscopy Facility (ETSF)}

\ead{richard.a.lynn@jyu.fi}

\begin{abstract}
We present an ongoing development of an existing code for calculating ground-state, steady-state, and transient properties of many-particle systems. The development involves the addition of the full four-index two electron integrals, which allows for the calculation of transport systems, as well as the extension to multi-level electronic systems, such as atomic and molecular systems and other applications. The necessary derivations are shown, along with some preliminary results and a summary of future plans for the code.
\end{abstract}

\section{Introduction}\label{sec:intro}
The development of molecular electronics has in recent years made it desirable to be able to simulate time-dependent many-body phenomena. However, computational limitations have a significant effect, with any realistic system being impossible to calculate with current computing power. Therefore, we are forced to adopt approximate techniques. These can include mean-field theories such as Hartree--Fock\cite{Hartree1928,Slater1930,Fock1930}, or methods such as density functional theory\cite{Hohenberg1964,Kohn1965,Runge1984,vanLeeuwen1999}, which rely on an approximate potential. We look here at non-equilibrium Green's function theory\cite{stefanuccibook}, which is in principle exact. However, it requires a truncation of an infinite sum term known as the self-energy.

When considering a many-particle system in the ground-state, it is well-known that properties can be calculated using many-body Green's functions, containing a suitable approximation for the self-energy, $\Sigma$. However, in order to study time-dependent properties of these systems, it is necessary to propagate these Green's functions in time by solving their equations of motion. The equations of motion for the many-particle Green's functions are given by the coupled integro-differential Kadanoff--Baym equations\cite{kadanoffbook,schafer1996,bonitz1996,binder1997}. The solution to these equations was calculated numerically by Dahlen and van Leeuwen in 2007\cite{dahlen2007}, leading to the analysis of many time-dependent systems, including inhomogeneous systems\cite{stan2009}, and strongly correlated few-electron quantum dots\cite{balzer2009}.

In 2009, this was extended to consider an embedded system, with a number of leads coupled to a central system of interest\cite{myohanen2009}. When propagated through time, both the transient and steady-state properties could be investigated, as well as the initial ground-state properties. This allowed for further investigations, such as image charge dynamics\cite{myohanen2012}, quantum transport in AC and DC fields\cite{myohanen2010}, and real-time switching between different steady states\cite{uimonen2010}.

In these earlier investigations, it was possible to use a simplification of the interaction term in the Hamiltonian, allowing for much faster computation. However, it has become desirable to include a fully described interaction, for the purposes of investigating transport and other applications using four-index two electron integrals, rather than the simplified two-index integrals used so far. The theory behind this expansion and the initial results and plans for the code in its new form are presented here.

\section{Theory}\label{sec:theory}

\subsection{Quantum Transport Model}\label{subsec:model}

When considering a two terminal quantum transport model, with a central system of interest connected to leads, the model Hamiltonian is described by:
\begin{equation}\label{eq:ham}
\hat{H}(t)=\hat{H}_C(t)+\hat{H}_{\textrm{leads}}(t)+\hat{H}_T-\mu\hat{N}
\end{equation}
where the Hamiltonian of the central system, $\hat{H}_C$, the Hamiltonians of the leads, $\hat{H}_{leads}$, and the tunneling Hamiltonian, $\hat{H}_T$, are given by Equation \ref{eq:hams}:
\begin{subequations}
\label{eq:hams}
\begin{align}
\hat{H}_C&=\sum_{ij}\sum_{\sigma}h_{ij}(t)\hat{a}_{i\sigma}^{\dagger}\hat{a}_{j\sigma}+\frac{1}{2}\sum_{ijkl}\sum_{\sigma\sigma^{\prime}}w_{ijkl}\hat{a}_{i\sigma}^{\dagger}\hat{a}_{j\sigma^{\prime}}^{\dagger}\hat{a}_{k\sigma^{\prime}}\hat{a}_{l\sigma}\label{subeq:hamc}\\
\hat{H}_{\textrm{leads}}(t)&=\sum_{\alpha=L,R}\sum_{\substack{i\in C\\j\in \alpha}}^{\infty}\sum_{\sigma}\left[h_{ij}^{\alpha}+W^{\alpha}(t)\delta_{ij}\right]\hat{c}_{i\sigma\alpha}^{\dagger}\hat{c}_{j\sigma\alpha}\label{subeq:haml}\\
\hat{H}_T&=\sum_{\alpha=L,R}\sum_{\substack{i\in C\\j\in \alpha}}\sum_{\sigma}V_{i,j\alpha}\left[\hat{d}_{i\sigma}^{\dagger}\hat{c}_{j\sigma\alpha}+\hat{c}_{j\sigma\alpha}^{\dagger}\hat{d}_{i\sigma}\right]\label{subeq:hamt}
\end{align}
\end{subequations}

In Equation \ref{subeq:haml} the Hamiltonian for the leads, $\alpha=L,R$ is described, with $h_{ij}^{\alpha}$ being the nearest neighbour Hamiltonian, $\hat{c}_{i\sigma\alpha}^{\dagger}$ and $\hat{c}_{j\sigma\alpha}$ are the creation and annihilation operators respectively for the lead, and $W^{\alpha}(t)$ is the local potential dependent on time. Typically this local potential is modelled as a bias voltage generated by applying an electric field over the system.

Equation \ref{subeq:hamt} describes the couplings between the central system and the leads, with $V_{i,j\alpha}$ being the matrix elements of the coupling Hamiltonian. The final term in Equation \ref{eq:ham} couples the chemical potential, $\mu$, to the total particle number operator, $\hat{N}$.

Of particular interest to this study are the one-body and two-body parts of Equation \ref{subeq:hamc}, $h_{ij}$ and $w_{ijkl}$. In previous studies, typically only the main local and non-local terms of these two terms were considered. In particular, this leads to the reduction of the interaction term $w_{ijkl}=\delta_{il}\delta_{kl}w_{ij}$, where $w_{ij}=1/\left|i-j\right|$ is the long-range behaviour of the matrix elements of the interaction. However, it has now become desirable to remove this assumption, using the full descriptions as given by:
\begin{subequations}
	\label{eq:ints}
	\begin{align}
	h_{ij}&=-\frac{1}{2}\int d\textbf{x}\phi_i^*(\textbf{x})\hat{h}(\textbf{x},t)\phi_j(\textbf{x})\label{subeq:hint}\\
	w_{ijkl}&=\int d\textbf{x}d\textbf{y}\phi_i^*(\textbf{x})\phi_j^*(\textbf{y})\hat{w}(\textbf{x},\textbf{y})\phi_k(\textbf{y})\phi_l(\textbf{x})\label{subeq:wint}
	\end{align}
\end{subequations}
where the standard position-spin coordinate description, $\textbf{x}=(\textbf{r},\sigma)$, has been used.

\subsection{The Kadanoff-Baym Equations}\label{subsec:KB}

\begin{figure}
	\centering
	\includegraphics[width=0.4\textwidth]{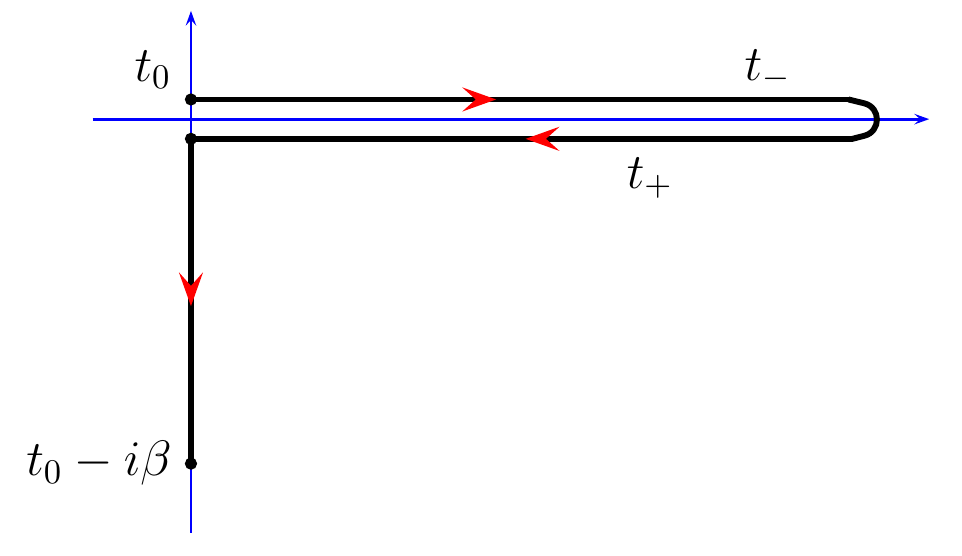}
	\includegraphics[width=0.4\textwidth]{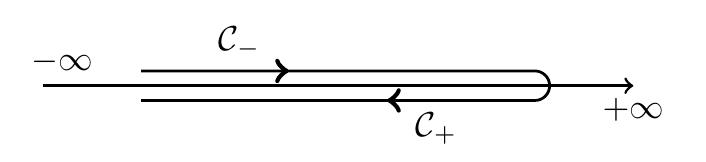}
	\caption{Keldysh Contours. The left gives the Keldysh contour showing the initial correlations in grand canonical ensemble form. The right gives a closed time-loop contour, $\mathcal{C}$. The forward time-ordered branch and backward anti-time-ordered branch are denoted by ``$-$'' and ``$+$'' respectively.}\label{fig:keldysh}
\end{figure}

For the purposes of transport calculations, the system is initially assumed to be in thermal equilibrium, before an external perturbation is applied at time $t_0$, after which time-evolution is carried out. The whole process is described by the Keldysh contour, as seen in Figure \ref{fig:keldysh}, with the ground-state described by the vertical imaginary track, and the dynamics described by the time-loop contour. Starting from the equations of motion for the single particle Green's function:
\begin{subequations}
	\label{eq:eom}
	\begin{align}
	\left[i\partial_{z_1}-\hat{h}(1)\right]G(1,2)&=\delta(1,2)+\int d3\Sigma_{\text{MB}}(1,3)G(3,2)\\
	G(1,2)\left[-i\overset{\leftarrow}{\partial}_{z_2}-\hat{h}(2)\right]&=\delta(1,2)+\int d3G(1,3)\bar{\Sigma}_{\text{MB}}(3,2)
	\end{align}
\end{subequations} 
different components can be distinguished on the contour $\mathcal{C}$ by utilising the Langreth rules. These components are called the Kadanoff-Baym equations, and are given as:
\begin{subequations}
	\label{eq:kb}
	\begin{align}
	\left[i\partial_{t_1}-\hat{h}(t_1)\right]G^{\gtrless}(t_1,t_2)&=\left[\Sigma^R\cdot G^{\gtrless}+\Sigma^{\gtrless}\cdot G^A+\Sigma^{\rceil}\star G^{\lceil}\right](t_1,t_2)=I_1^{\gtrless}(t_1,t_2)\\
	G^{\gtrless}(t_1,t_2)\left[-i\overset{\leftarrow}{\partial}_{t_2}-\hat{h}(t_2)\right]&=\left[G^R\cdot\Sigma^{\gtrless}+G^{\gtrless}\cdot\Sigma^A+G^{\rceil}\star\Sigma^{\lceil}\right](t_1,t_2)=I_2^{\gtrless}(t_1,t_2)\\
	\left[i\partial_{t_1}-\hat{h}(t_1)\right]G^{\rceil}(t_1,\tau_2)&=\left[\Sigma^R\cdot G^{\rceil}+\Sigma^{\rceil}\star G^M\right](t_1,\tau_2)=I^{\rceil}(t_1,\tau_2)\\
	G^{\lceil}(\tau_1,t_2)\left[-i\overset{\leftarrow}{\partial}_{t_2}-\hat{h}(t_2)\right]&=\left[G^{\lceil}\cdot\Sigma^A+G^M\star\Sigma^{\lceil}\right](\tau_1,t_2)=I^{\lceil}(\tau_1,t_2)\\
	\left[-\partial_{\tau_1}-\hat{h}^M\right]G^M(\tau_1,\tau_2)&=i\delta(\tau_1-\tau_2)+\left[\Sigma^M\star G^M\right](\tau_1,\tau_2)=I^M(\tau_1,\tau_2)\\
	G^M(\tau_1,\tau_2)\left[\delta_{\tau_2}-\hat{h}^M\right]&=i\delta(\tau_1-\tau_2)+\left[G^M\star\Sigma^M\right](\tau_1,\tau_2)=I^M(\tau_1,\tau_2)
	\end{align}
\end{subequations}
The short-hand notation, $I^{\gtrless,\rceil,\lceil}$, refers to collision integrals, and the convolution integrals are defined as:
\begin{subequations}
	\label{eq:conv}
	\begin{align}
	\left[a\cdot b\right](t_1,t_2)&=\int_{t_0}^{\infty}a(t_1,t)b(t,t_2)dt\\
	\left[a\star b\right](t_1,t_2)&=-\int_{0}^{\beta}a(t_1,\tau)b(\tau,t_2)d\tau
	\end{align}
\end{subequations}

Solving the Kadanoff--Baym equations numerically is far from trivial. The initial step is to solve the system along the imaginary time axis, that is the Matsubara component of the Green's function. The equation of motion of the Matsubara component is written as:
\begin{equation}
\sum_k\left(-\delta_{ik}\partial_{\tau_1}-h_{ik}^0\right)G_{kj}^M(\tau_1)=\delta_{ij}\delta(\tau_1)+\int_{0}^{\beta}d\tau_3\sum_k\Sigma_{ik}^M(\tau_1-\tau_3)G_{kj}^M(\tau_3)\label{eq:matseom}
\end{equation}
Equation \ref{eq:matseom} can then be cast in the form of a Dyson equation:
\begin{equation}
G_{ij}^M(\tau_1)=G_{0,ij}(\tau_1)+\int_{-\beta}^{0}d\tau_4\int_{0}^{\beta}d\tau_3\sum_{kl}G_{0,ik}(\tau_1-(\tau_3-\beta))\Sigma_{kl}^{M,c}(\tau_3-(\tau_4+\beta))G_{lj}^M(\tau_4)\label{eq:matsdys}
\end{equation}
which can be solved iteratively to self-consistency using, e.g., Hartree--Fock. Using the solution to Equation \ref{eq:matsdys}, the two-time propagation can be started. By making use of symmetries, the subset of equations requiring propagation can be reduced to:
\begin{subequations}
	\label{eq:prop}
	\begin{align}
	i\partial_{t_1}G^{\textgreater}(t_1,t_2)&=\hat{h}(t_1)G^{\textgreater}(t_1,t_2)+I_1^{\textgreater}(t_1,t_2)\\
	-i\partial_{t_1}G^{\textless}(t_2,t_1)&=G^{\textless}(t_2,t_1)\hat{h}(t_1)+I_2^{\textless}(t_2,t_1)\\
	i\partial_{t_1}G^{\rceil}(t_1,-i\tau)&=\hat{h}(t_1)G^{\rceil}(t_1,-i\tau)+I^{\rceil}(t_1,-i\tau)\\
	-i\partial_{t_1}G^{\lceil}(-i\tau,t_1)&=G^{\lceil}(-i\tau,t_1)\hat{h}(t_1)+I^{\lceil}(-i\tau,t_1)
	\end{align}
\end{subequations}
These equations can finally be propagated by using a Hartree--Fock Hamiltonian with small time-steps $\Delta$ from $t\rightarrow t+\Delta$.

\subsection{Self Energy Approximations}\label{subsec:sigma}

\begin{figure}
\begin{center}
	\begin{tabular}{m{25pt}m{80pt}m{15pt}m{100pt}}
		$\Sigma^{\text{HF}}=$
		&
		\begin{fmffile}{HSigma}
			\begin{fmfgraph*}(100,70)
				\fmfleft{i}
				\fmfright{o}
				\fmfbottom{b}
				\fmftop{t}
				\fmf{photon,tension=0.2}{b,v1}
				\fmf{phantom,tension=5}{v2,t}
				\fmf{fermion,left,tension=0.1}{v1,v2,v1}
			\end{fmfgraph*}
		\end{fmffile}
		&
		+
		&
		\begin{fmffile}{FSigma}
			\begin{fmfgraph*}(80,70)
				\fmfleft{i}
				\fmfright{o}
				\fmf{fermion}{o,i}
				\fmf{photon,left,tension=0.5}{i,o}
			\end{fmfgraph*}
		\end{fmffile}
	\end{tabular}
\end{center}

\begin{center}
	\begin{tabular}{m{75pt}m{100pt}m{5pt}m{100pt}}
		$\Sigma^{(2)}=\Sigma^{\text{HF}}+$
		&
		\begin{fmffile}{2Sigma}
			\begin{fmfgraph*}(100,75)
				\fmfleft{i}
				\fmfright{l}
				\fmf{fermion}{l,v2}
				\fmf{fermion}{v2,v1}
				\fmf{fermion}{v1,i}
				\fmf{photon,left,tension=0.1}{i,v2}
				\fmf{photon,right,tension=0.1}{v1,l}
			\end{fmfgraph*}
		\end{fmffile}
		&
		+
		&
		\begin{fmffile}{BSigma}
			\begin{fmfgraph*}(100,50)
				\fmfleft{i,j}
				\fmfright{k,l}
				\fmf{fermion}{k,i}
				\fmf{fermion,right=.5}{j,l}
				\fmf{electron,right=.5}{l,j}
				\fmf{photon}{i,j}
				\fmf{photon}{k,l}
			\end{fmfgraph*}
		\end{fmffile}
	\end{tabular}
\end{center}
	\caption{\label{fig:sigma}Diagrammatic representations of the self-energy approximations. The Hartree--Fock approximation, top, consists of the Hartree term, and the Fock, or exchange term. The second-order Born approximation consists of the Hartree--Fock self-energy, added to the first order bubble diagram, and the second-order correction to the exchange term.}
\end{figure}
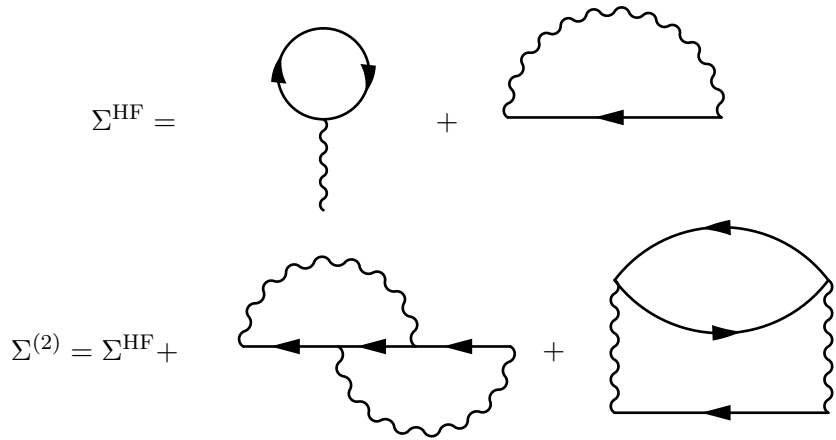

The most important approximation used in the theory is that of the self-energy. This term is expanded in terms of the interaction, and we here consider two approximations. The first is the well-known Hartree--Fock approximation, described by Equation \ref{eq:HF}. The first term in this is the Hartree term, and the second is the Fock, or exchange, term. The other expansion used is the second-order Born approximation, given by Equation \ref{eq:2B}.\footnote{The second-order Born self-energy is not directly used in this work, but when time-dependence is introduced, it is intended that it shall be utilised substantially.} The first of the terms beyond Hartree--Fock is commonly referred to as the first order bubble term, while the final term is simply the second order correction to the exchange term. Both of these self-energies are presented diagrammatically in Figure \ref{fig:sigma}.

\begin{equation}\label{eq:HF}
\Sigma^{\text{HF}}_{ij}(t)=-i\sum_{kl}G_{kl}(t,t^{+})(2w_{ilkj}-w_{iljk})
\end{equation}

\begin{equation}\label{eq:2B}
\Sigma_{ij}^{(2)}(t,t^{\prime})=\sum_{klmnpq}G_{kl}(t,t^{\prime})G_{mn}(t,t^{\prime})G_{pq}(t,t^{\prime})\times w_{iqmk}(2w_{lnpj}-w_{nlpj})
\end{equation}

\subsection{Basis Sets}\label{subsec:basis}

The one- and two-body parts of the Hamiltonian, described by Equation \ref{eq:ints}, depend on a chosen basis set. Any basis set can be used for this purpose, but for this current investigation, an atom-centred Slater basis set has been used. The definition of a Slater type orbital is given by:
\begin{equation}\label{eq:sto}
S_{nlm}^{\zeta}(r,\theta,\phi)=Nr^{n-1}e^{-\zeta r}Y_l^m(\theta,\phi)
\end{equation}
where $n$, $l$, and $m$ are the principle quantum numbers, which follow the standard rule $n\textgreater l\geq|m|\geq 0$. $N$ is a normalisation constant, and $Y_l^m$ are spherical harmonics, given by:
\begin{subequations}\label{eq:spher}
\begin{align}
Y_l^m(\theta,\phi)&=Ne^{im\phi}P_l^m(\cos\theta)\\
P_l^m(x)&=\frac{(-1)^m}{2^ll!}(1-x^2)^{m/2}\frac{d^{l+m}}{dx^{l+m}}(x^2-1)^l
\end{align}
\end{subequations} 
where $P_l^m$ are the associated Legendre polynomials.

As the Slater type orbitals are based on the eigenvalues of the Hydrogen atom, they are beneficial in reducing the time taken to convergence in the calculations of molecular orbitals. When calculating these bases, it is possible to use ready prepared tables with values of $\zeta$ for all combinations of $n$, $l$, and $m$ for different atoms. Using a seperate basis set for each atom, the Hartree--Fock orbitals are found by solving a generalised eigenvalue problem, and these orbitals then form a new orthonormal basis set.\footnote{For more details, the full basis sets and a file of Slater functions and exponents are available for the reader upon request.}

\section{Initial Results}\label{sec:results}

While the complete implementation has yet to be completed, there have already been some preliminary ground-state results obtained as a test of the reliability of the code. These initial tests were carried out on the hydrides of the first row elements, Lithium to Fluorine. These molecules were chosen as exact theoretical values for many of their properties are known, so can be directly compared. The size of the basis set used for the calculations was dependent on the atoms involved, i.e. there was one basis function for all different values of $n$, $l$, and $m$ for each of the two atoms. 

\begin{figure}
	\centering
	\includegraphics[width=0.8\textwidth]{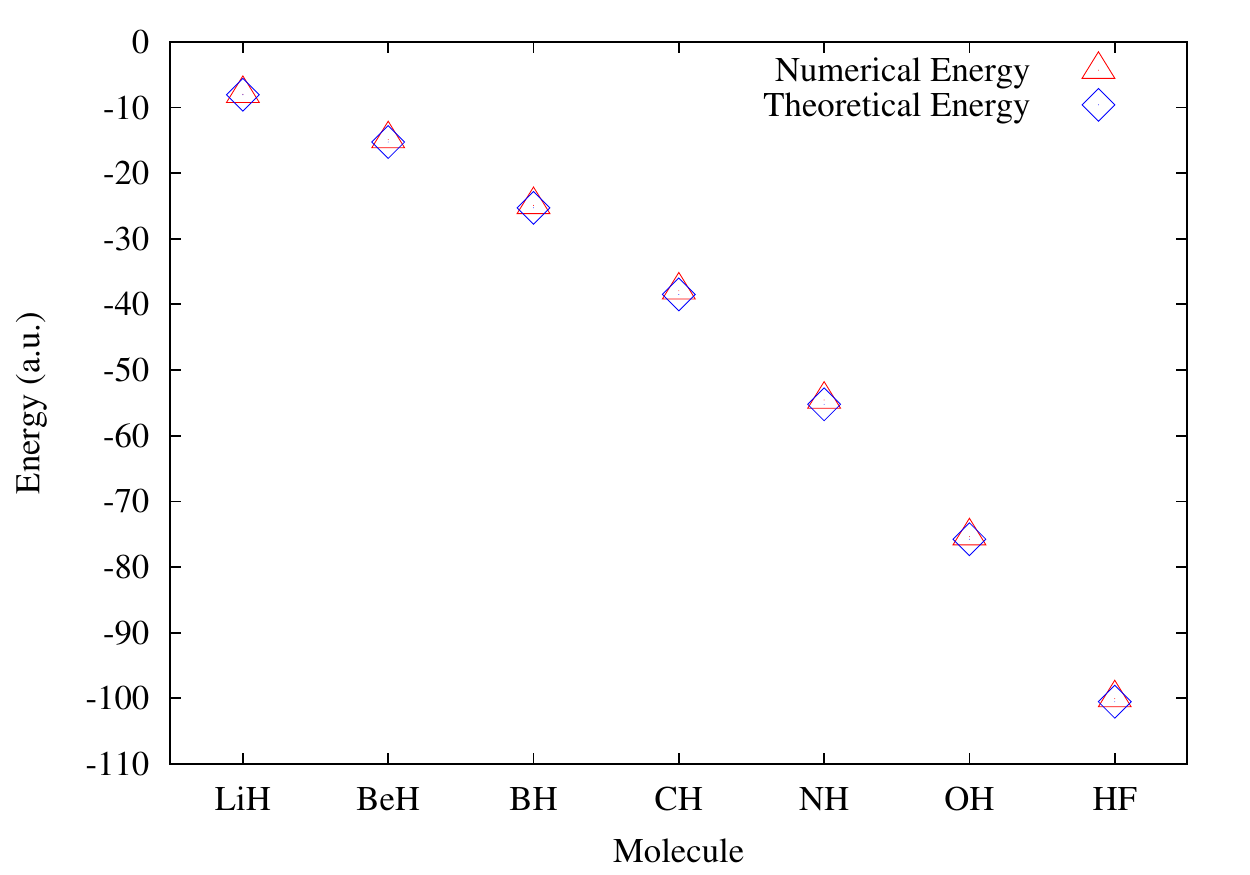}
	\caption{\label{fig:gsenerg}The ground-state energies of diatomic molecules calculated at the equilibrium inter-atomic bond length. The theoretical results are shown as blue diamonds, while the Hartree--Fock results are displayed as red triangles.}
\end{figure}

The first test was to find the ground-state equilibrium energy of the molecule. The theoretical values for comparison were calculated by summing the individual ground-state energies of the individual atoms, and adding the experimental dissociation energy. The numerical results were achieved by finding the ground-state energy of the molecule at the experimental equilibrium bond distance. Table \ref{tab:gsenerg} shows the two energies, and they are plotted for comparison in Figure \ref{fig:gsenerg}. It can be seen that the agreement is very good between the two figures, with the numerical results slightly overestimating the equilibrium energy, as would be expected.

\begin{table}
	\centering
	\begin{tabular}{@{}llll@{}}
		\toprule
		\bfseries{Molecule}&\bfseries{Equilibrium Bond Length}&\bfseries{Theoretical Energy}&\bfseries{Numerical Energy}\\
		\midrule
		LiH&3.0094&-8.0701&-7.984791\\
		BeH&2.5340&-15.2529&-14.881285\\
		BH&2.2453&-25.2815&-24.920593\\
		CH&2.0566&-38.4817&-37.916774\\
		NH&1.9057&-55.2283&-54.555561\\
		OH&1.8113&-75.7694&-75.385111\\
		HF&1.7302&-100.5197&-100.063868\\
		\bottomrule
	\end{tabular}		
\caption{The equilibrium bond lengths of diatomic molecules, together with the theoretical and numerical ground-state energy. All figures are in atomic units.}\label{tab:gsenerg}
\end{table}

\begin{table}
	\centering
	\begin{tabular}{@{}lllll@{}}
		\toprule
		\bfseries{Molecule}&\bfseries{Experimental $D_e$}&\bfseries{Morse $D_e$}&\bfseries{Rydberg $D_e$}&\bfseries{Varshni $D_e$}\\
		\midrule
		LiH&0.0926&0.0839041&0.0751182&0.0417132\\
		BeH&0.0847&0.0462033&0.0417274&0.0298187\\
		BH&0.1243&0.224196&0.204235&0.123077\\
		CH&0.1270&0.313989&0.281394&0.138535\\
		NH&0.1182&0.0182241&0.0164693&0.0142616\\
		OH&0.1616&0.110868&0.106774&0.0979529\\
		HF&0.2151&0.375695&0.340266&0.18733\\
		\bottomrule
	\end{tabular}		
	\caption{The experimental dissociation energies and the energies calculated by fitting data to the Morse, Rydberg, and Varshni inter-atomic potentials.}\label{tab:dissenerg}
\end{table}

\begin{figure}
	\centering
	\includegraphics[width=0.9\textwidth]{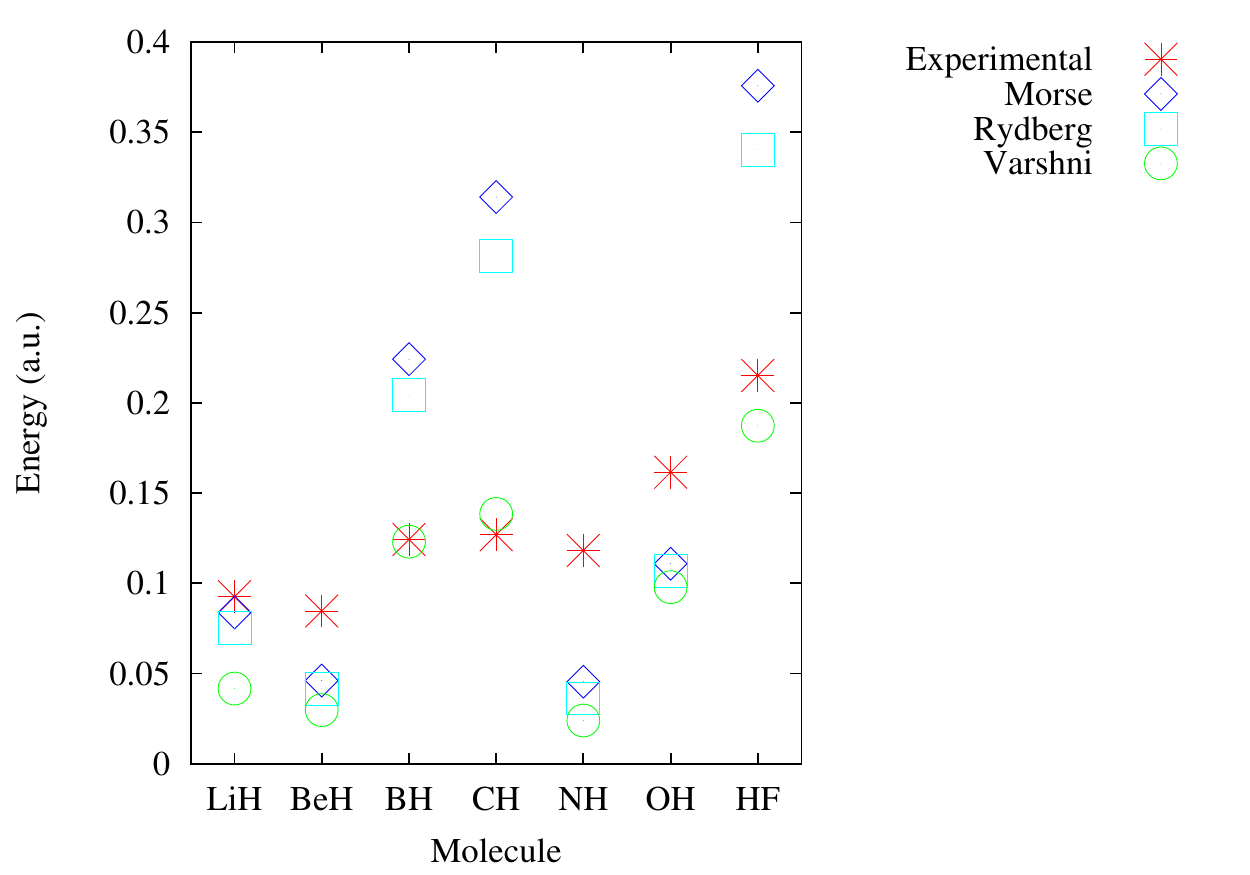}
	\caption{\label{fig:dissenerg}The dissociation energies of diatomic molecules calculated by fitting three different potentials to numerical data. The experimental energies are shown as red stars, blue diamonds show the Morse potential fit results, cyan squares show the results from the Rydberg potential, and the green circles show the Varshni potential.}
\end{figure}

\begin{figure}
	\centering
	\includegraphics[width=0.9\textwidth]{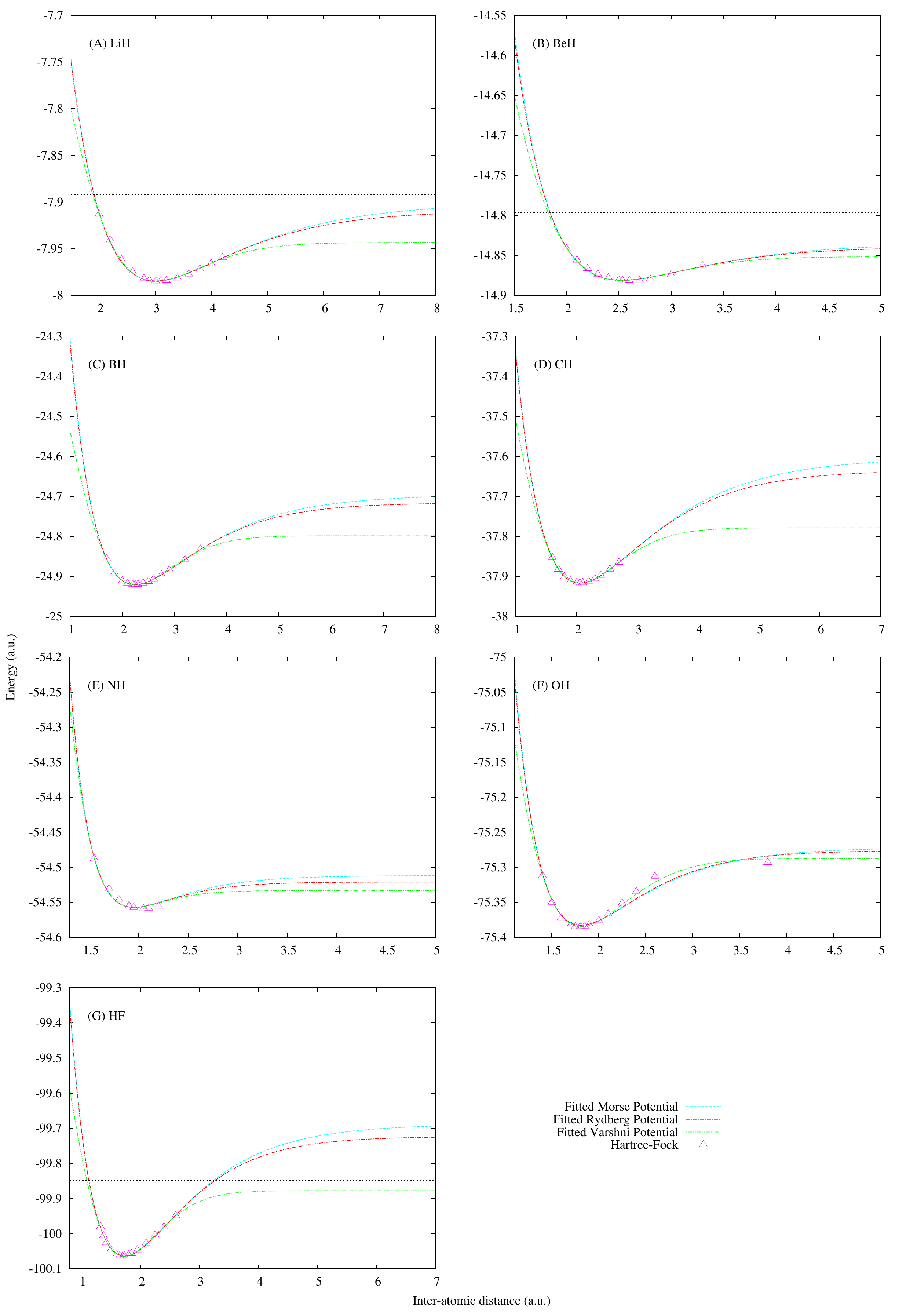}
	\caption{\label{fig:disscurves}The dissociation curves for the first row hydrides. The Hartree--Fock results for various inter-atomic distances are shown as purple triangles, and various approximate atomic potentials are fitted to the data. A Morse potential is shown as a dot-dash-dotted green line, Rydberg potential as a dot-dashed red line, and Varshni potential as a dashed cyan line. The black dashed line shows the experimental dissociation energy of the molecule for comparison.}
\end{figure}

Having found the ground-state equilibrium results, the other significant property that was investigated was the dissociation energy. This was found by finding the numerical energy for various different bond lengths, and then fitting a common inter-atomic potential curve to the data. There were three potentials used to provide a thorough test, as well as give a range of energies. The first potential used was the Morse potential, given by:
\begin{equation}
V(r)=E+D_e\left(1-e^{-b(r-r_e)}\right)^2
\label{eq:morse}
\end{equation}
with fitting parameters $E$, $D_e$, and $b$. The second potential was the Rydberg potential, given by:
\begin{equation}
V(r)=E+D_e\left(1-\left(1+\frac{k}{D_e}\right)^{\frac{1}{2}}(r-r_e)\right)e^{-\left(\frac{k}{D_e}\right)^{\frac{1}{2}}(r-r_e)}
\label{eq:rydberg}
\end{equation}
with fitting parameters $E$, $D_e$, and $k$. The final potential used was the Varshni potential, given by:
\begin{equation}
V(r)=E+D_e\left(1-e^{-b\left(r^2-r_e^2\right)}\right)^2
\label{eq:varshni}
\end{equation}
with fitting parameters $E$, $D_e$, and $b$. For all three potentials, the value of the fitting parameter $D_e$ gives the approximate dissociation energy. The values of $D_e$ for each potential and molecule are shown in Table \ref{tab:dissenerg} with the experimental values for comparison, and the results are plotted in Figure \ref{fig:dissenerg}. The individual dissociation curves for each molecule are shown in Figure \ref{fig:disscurves}. 

The results shown here suggest that the calculation of dissociation energy is less accurate than the equilibrium energy value. However, this is in part due to the limited amount of data that has been used to fit the potential curves, and an inherent inaccuracy within the potentials themselves. The Varshni potential is the most consistent approximation, and follows roughly the correct trend. As these results were obtained primarily as a test of the functionality of the code, it can be said that the relative proximity between experimental and numerical results prove that the code is working as expected.

\section{Conclusion and Future Plans}\label{sec:concl}
The initial results presented here show that the development of the code has already been successful for the calculation of ground-state properties, and implementation and testing of the time propagation have already been started for finding transient and steady-state properties. Once this has been completed, several specific applications have been planned for the future, which shall be briefly detailed here.

The topic of shot noise in quantum systems is an active area of research, and the more complex form of the interactions between particles being used in this new implementation make it possible to simulate. The system itself is a typical transport setup, with a central system of interest between two leads. Initially, the current over this system is measured as a function of time. If this is plotted, a fluctuation around the mean current can be seen. It is known that these fluctuations are correlated, and the current-current correlation function is defined as $\Delta I_{\alpha}(t,t^{\prime})=I_{\alpha}(t)-\left\langle I_{\alpha}(t)\right\rangle$, where the current operator is defined as the time derivative of the total number of particles, given by $N_{\alpha}(t)=\sum_kc^{\dagger}_{k\alpha}(t)c_{k\alpha}(t)$. The property of interest is the time-dependent shot noise, which is defined as:
\begin{equation}
S_{\alpha\beta}(t,t^{\prime})\equiv\frac{1}{2}\left\langle\Delta I_{\alpha}(t)\Delta I_{\beta}(t^{\prime})+\Delta I_{\beta}(t^{\prime})\Delta I_{\alpha}(t)\right\rangle
\label{eq:shotnoise}
\end{equation}

The other investigation planned using the new implementation is with the well-known Anderson impurity model. Typically, the simplified interaction used in the Anderson model removes the possibility of multiple energy-levels on a single site. By using the full interaction term, it is hoped that more can be understood about the simple one-site model, as well as providing the first stage in more complex systems. An isolated central site will first be studied, before the contacted case is considered and compared. It would be expected that a broadening of the peaks from the isolated case would be observed, along with the addition of a Kondo peak, as in the simplified model, but it is possible that additional physics may be observed.

\ack
We would like to acknowledge the Finnish Academy of Science for providing the funding for this project to be carried out.

\appendix

\section{Numerical Implementation}\label{app:details}

The calculation of the two-electron integrals, given by Equation \ref{subeq:wint}, makes up the majority of the additional code in this new implementation. As the integrand is made up of a product of four basis functions, with form described by Equations \ref{eq:sto} and \ref{eq:spher}, along with the interaction $w(\textbf{x},\textbf{y})$, this is a non-trivial integral to solve.

Before the integration itself, the variables within the bases are first read in from an input file. At this stage the input provides capability for modelling single atoms and diatomic molecules, but this can later be expanded to an $N$-particle system. As an initial simplification, those integrals where all four bases are centred on the same atomic site are considered separately.

The approach used considers the integral in spherical polar coordinates, with the separation of variables where possible simplifying the process considerably. The resulting multiple integrals are solved using a set of different numerical integration techniques.

The first main method used is Gauss-Legendre quadrature. This effectively reduces an integral to a sum of weightings applied to the function being integrated at fixed points. In general in Gaussian quadrature, the fixed points are determined by the roots of an associated polynomial. In the case of Gauss-Legendre integration, the associated polynomials are Legendre polynomials. This method is applied to integrands such as:
\begin{equation}
\int_1^{\infty}dxx^j\left(x^2-1\right)^me^{-cx}
\label{eq:gl1}
\end{equation}
and
\begin{equation}
\int_{-1}^{1}dxx^j\left(1-x^2\right)e^{-cx}
\label{eq:gl2}
\end{equation}

The other numerical method used significantly is downwards recursion. This method can be applied, e.g. when the solution to an integral can depend on the solution to another integral with the same form of the integrand, but lower polynomial powers. This is applied in this code to situations such as:
\begin{equation}
\int_1^{\infty}dxx^je^{-cx}
\label{eq:recur1}
\end{equation}
and
\begin{equation}
\frac{-1^l}{2}\int_{-1}^{1}dtP_l(t)e^{-\beta t}
\label{eq:recur2}
\end{equation}

In addition to this, there are some integrals with simple known solutions, which are calculated directly, before the final value of $w_{ijkl}$ is formed by combining these results.

For storing the two electron integrals, they must be kept in a two-dimensional array. As they are a four-dimensional object, this means some adjustment of the indices. In practice, they are stored as $v[p][q]$, where $i=1+(p-1)/nb$ and $j=p-nb*(i-1)$, with similar relations for $k$,$l$,and $q$.


\bibliography{RALandRVL2015}

\end{document}